\documentclass[twoside,a4paper,aps,prl,floatfix,twocolumn,showpacs,superscriptaddress,10pt,nobibnotes]{revtex4-1}

\usepackage{graphicx}
\usepackage{color}
\usepackage{tabularx}
\usepackage{epsfig}
\usepackage{amsmath}
\usepackage{amssymb}
\usepackage{amsfonts}
\usepackage{graphicx}
\usepackage{dcolumn}
\usepackage{bm}
\usepackage{wasysym}
\usepackage{times}
\usepackage{latexsym}

\definecolor{grn}{rgb}{0,0,0.54}

\newcommand{\bra}[1]{\ensuremath{\langle #1 |}}
\newcommand{\ket}[1]{\ensuremath{| #1 \rangle}}

\begin{document}

\title{Coexistence of long-range and algebraic correlations for short-range valence-bond wave functions in three dimensions}

\author{A. Fabricio Albuquerque}
\affiliation{Instituto de F{\' \i}sica, Universidade Federal do Rio de Janeiro, Cx.P.~68.528, 21941-972 Rio de Janeiro-RJ, Brazil}
\affiliation{Laboratoire de Physique Th{\' e}orique, Universit{\' e} de Toulouse and CNRS, UPS (IRSAMC),
F-31062 Toulouse, France}
\author{Fabien Alet}
\affiliation{Laboratoire de Physique Th{\' e}orique, Universit{\' e} de Toulouse and CNRS, UPS (IRSAMC),
F-31062 Toulouse, France}
\author{R. Moessner}
\affiliation{Max-Planck-Institut f\"{u}r Physik komplexer Systeme, 01187 Dresden, Germany}

\date{\today}


\begin{abstract}
We investigate nearest-neighbor valence-bond wave functions on bipartite three-dimensional lattices. By performing large-scale Monte Carlo simulations, we find that
long-range magnetic order coexists with dipolar four-spin correlations on the cubic lattice, this latter feature being reminiscent of the Coulomb phase for classical dimers
on the same geometry. Similar properties are found for the lower-coordination diamond lattice. While this suggests that the coexistence of magnetic order and dipolar four-spin
correlations is generic for such states on bipartite three-dimensional lattices, we show that simple generalizations of these wave functions can encode different ordering behaviors.
\end{abstract}

\maketitle


Given the difficulty to obtain ground states for even simple model Hamiltonians, one approach in trying to find novel strongly correlated states is to study variational or trial wave functions directly, and to seek the Hamiltonians which they minimize in a subsequent step. Most prominently, such an approach has been successfully pursued in the case of Laughlin's wave function for the fractional quantum Hall effect \cite{laughlin:83}.

For unusual magnetic states, wave functions defined in terms of valence-bonds (VBs) have an illustrious history \cite{liang:88,baskaran:87}. In their simplest incarnation, in the context of quantum dimer models, a dimer represents a VB. This has established a connection between the venerable subject of close-packed hard-core dimer coverings \cite{kasteleyn:61,temperley:61,fisher:61,fisher:63} and the physics of strongly correlated electrons, and has led to important conceptual progresses, including the discovery of a liquid phase with exponentially decaying correlations \cite{moessner:01}. More recently, the problem of dimers on the cubic and other {\em bipartite} three-dimensional (3d) lattices has received considerable attention \cite{huse:03,hermele:04}. The hard-core constraint for dimers allows for the formulation of an effective theory in terms of a coarse-grained, divergence-less, magnetic field \cite{huse:03}, implying that dimers on bipartite 3d lattices realize a Coulomb phase \cite{henley:10}, characterized by dipolar dimer-dimer correlations. We also briefly mention the close relationship \cite{isakov:04,hermele:04} between hard-core dimers on the diamond lattice
and {\em spin-ice} systems on the pyrochlore lattice \cite{gardner:10}, that have been the focus of intensive research in recent years \cite{bramwell:01,castelnovo:12}.

VBs or related degrees of freedom also appear in several other fields of research, including quantum chemistry, from where they originated (see e.g.~Refs.~\onlinecite{rumer:32,
rumer:32b}). The intimate relationship between superimposed VB configurations and loop models was later established, the latter being intensively studied in statistical
mechanics \cite{nienhuis:10}, for e.g.~their geometric transitions~\cite{nahum:11,nahum:12,jaubert:11}. In quantum information, entanglement properties of short-range
VB wave functions have been recently studied~\cite{verstraete:06,poilblanc:12,schuch:12}. Finally, we mention the recent experimental realization of a plaquette
VB state with ultracold atoms loaded into optical superlattices~\cite{nascimbene:12}.

\begin{figure}
\begin{center}
  \includegraphics*[width=0.215\textwidth,angle=270]{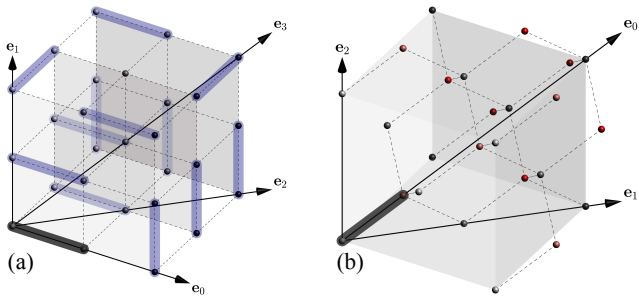}
  \caption{
    (Color online) (a) Cubic and (b) diamond lattices considered in the present work. Thick light-shaded links in (a)
                              represent singlets in a NN-VB state [Eq.~\eqref{eq:psi}]. We also indicate special directions (with respect to a reference dimer
                              depicted as a thick dark line), along which four-point correlations are computed.
                              }
  \label{fig:lattices}
  \end{center}
\end{figure}

Within the framework of quantum magnetism, we remark that hard-core dimer approaches, while leading to exact results even in the 3d case, do not
{\em a priori} incorporate actual spin degrees of freedom. However, it is also crucial to be able to account for magnetic order and its interplay with
phases lacking it, in order to obtain a full picture of the nature of possible states. In this work, we analyze the effects due to the incorporation of spin
degrees of freedom by characterizing nearest-neighbor VB (NN-VB) wave functions defined on the 3d bipartite simple cubic and diamond lattices (see
Fig.~\ref{fig:lattices}). Such states can be regarded as 3d generalizations of the NN-VB wave function on the square lattice originally introduced by Sutherland
\cite{sutherland:88}, that has been recently studied via Monte Carlo (MC) simulations \cite{albuquerque:10b,tang:11}.
NN-VB wave functions on the cubic lattice have been previously analyzed in Ref.~\cite{beach:07}, a combined mean-field and MC study of a family of
Hamiltonians with interactions of varying ranges for spins of varying lengths. This work already found the presence of magnetic order -- see below (we
also mention recent studies of NN-VB states on frustrated 2d lattices~\cite{wildeboer:12,yang:12,schuch:12}). Here we present, for  cubic and diamond
lattices, a thorough evaluation of the correlations -- e.g. the presence of {\em dipolar} four-spin correlators -- encoded by NN-VB states, characterize
the ensemble of loops encountered, and provide an outline of a field-theoretic description of an extended family of NN-VB wavefunctions.

We consider the NN-VB state defined as an equal-amplitude superposition of {\em all} hard-core dimer coverings $c$ on the considered lattice,
\begin{equation}
\ket{\psi}=\sum_{c} \ket{c}~,
\label{eq:psi}
\end{equation}
where a singlet state, $[i,j]=\frac{1}{\sqrt{2}}(\ket{\! \! \uparrow_{i} \downarrow_{j}}-\ket{\! \! \downarrow_{i} \uparrow_{j}})$, is attached to each dimer joining site $i$ on sublattice  ${\mathcal A}$ to one of its NN sites $j$ on sublattice  ${\mathcal B}$. This fixes the overall sign of the wave function, so that stochastic methods are applicable \cite{liang:88}. The intimate relationship between classical dimers and NN-VB states in this definition suggests that the Coulomb phase found for dimers \cite{huse:03} may also be encoded in $SU(2)$-invariant NN-VB wave-functions. It is our main goal here to test this conjecture, by performing state-of-the-art MC simulations. We find an intricate interplay between magnetic and VB (energy) correlations, most remarkably in the form of N{\'e}el order coexisting with dipolar four-spin correlations.


{\it Numerical method and observables --- }
A crucial aspect of the MC simulations is to take into account the non-orthogonality of VB configurations~\cite{sutherland:88}.
Following Ref.~\onlinecite{liang:88}, we write the expectation value of an observable ${\mathcal O}$ in the wave function Eq.~\eqref{eq:psi} as
\begin{equation}
\langle {\mathcal O} \rangle = \frac{1}{Z} \sum_{c_{1},c_{2}} \frac{\bra{c_{1}} {\cal O} \ket{c_{2}}}{\langle c_{1} | c_{2} \rangle} \langle c_{1} | c_{2} \rangle~.
\label{eq:aver}
\end{equation}
$Z= {\langle \psi | \psi \rangle}$ stands for the normalization and the overlap between NN-VB configurations $\ket{c_{1}}$ and $\ket{c_{2}}$ is given by
${\langle c_{1} | c_{2} \rangle} =2^{N_{\mathcal{L}}(1,2) - \frac{N}{2}}$, with $N_{\mathcal{L}}(1,2)$ denoting the number of loops in the transition graph for both
configurations and $N$ the number of sites \cite{sutherland:88}. Alternatively, the normalization can be written in a pure loop representation as~\cite{sutherland:88}
\begin{equation}
Z=\sum_{\cal L} 2^{n_{l=2}({\cal L})} 4^{n_{l>2}({\cal L})},
\label{eq:Zloop}
\end{equation}
where the sum is over the set of all loop coverings obtained by superposing two NN-VB configurations, and $n_{l}({\cal L})$ the number of loops of size $l$ in the loop covering $ {\cal L}$.
The estimator ${\bra{c_{1}} {\cal O} \ket{c_{2}}}/{\langle c_{1} | c_{2} \rangle}$ for ${\mathcal O}$ in Eq.~\eqref{eq:aver},
determined by the loop structure in the transition graphs (for instance, the squared staggered magnetization $m_{\rm S}^{2}$ relates to the squared loop-lengths;
see Ref.~\onlinecite{beach:06}), can thus be sampled by generating {\em pairs} of dimer configurations, by relying on an efficient worm MC algorithm \cite{sandvik:10}.
Following this procedure, we probe two- and four-spin correlation functions for NN-VB states on the simple cubic and diamond lattices, for system sizes of up to $L=64$
(cubic) and $L=32$ (diamond, $2 \times 32^{3}$ sites) with periodic boundary conditions. Note that we do not take the existence of topological
sectors~\cite{huse:03} explicitly into account. In contrast with the square lattice case~\cite{albuquerque:10b,tang:11}, {\em winding} worms occur with
non-negligible frequency during the simulations, and consequently different topological sectors are sampled. This property stems from the existence of long-range
antiferromagnetic (AF) order, as we discuss next.


\begin{figure}
\begin{center}
  \includegraphics*[width=0.45\textwidth]{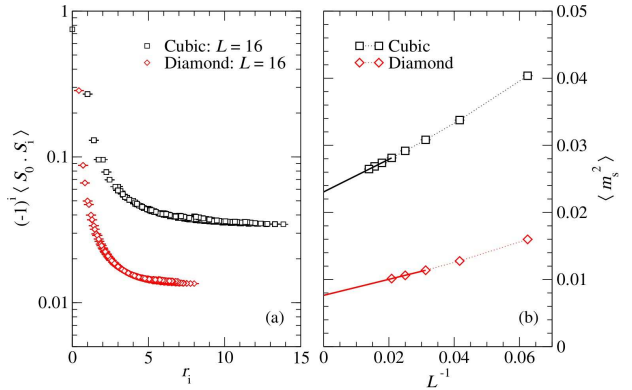}
  \caption{
    (Color online) (a) Two-spin staggered correlations as a function of distance $r$ for the NN-VB state on the cubic and diamond lattices (linear size is $L=16$). (b) Squared staggered magnetization density versus inverse system size
                                    $L^{-1}$. Extrapolated values, obtained from linear fits to the three left-most data points, are: $\langle m_{\rm S}^{2} \rangle_{\infty} = 0.02306(8)$
                                    (cubic) and $\langle m_{\rm S}^{2} \rangle_{\infty} = 0.0076(2)$ (diamond).
  }
  \label{fig:spin}
  \end{center}
\end{figure}

{\it Magnetic order --- }
We first consider staggered two-spin correlations, $(-1)^{\mathbf{r}} \langle {\mathbf{S}_{\mathbf{0}}} \cdot {\mathbf{S}_{\mathbf{r}}} \rangle$, as a function of distance
$r$ for cubic and diamond lattices with $L=16$. As shown in Fig.~\ref{fig:spin}(a), correlations seemingly saturate at large distances, suggesting that NN-VB wave
functions on both lattices display long-range N\'eel order. Previously obtained in Ref.~\onlinecite{beach:07} for the case of the cubic lattice (see also a related result for
higher spins \cite{parameswaran:09}), this under-appreciated
result is rather surprising if we recall that such wave functions were originally introduced as variational states for {\em magnetically disordered} systems. We put this
observation onto firmer grounds by analyzing the dependence on system size of the squared staggered moment, $\langle m_{\rm S}^{2} \rangle=\frac{1}{N^2}
\sum_{i,j} (-1)^{\mathbf{r_{ij}}} \langle {\mathbf{S}_{\mathbf{i}}} \cdot {\mathbf{S}_{\mathbf{j}}} \rangle$. Standard finite-size scaling arguments \cite{neuberger:89}
predict $\langle m_{\rm S}^{2} \rangle$ to scale as $L^{-1}$ for AF ordered states, in good agreement with our plot Fig.~\ref{fig:spin}(b). Data extrapolation confirms
magnetic ordering, with extrapolated values $\langle m_{\rm S}^{2} \rangle_{\infty} =
0.02306(8)$ (cubic) and $\langle m_{\rm S}^{2} \rangle_{\infty} = 0.0076(2)$ (diamond). While the lower coordination of the diamond lattice weakens AF order, it is
not enough to destabilize it. 

\begin{figure}
\begin{center}
  \includegraphics*[width=0.45\textwidth]{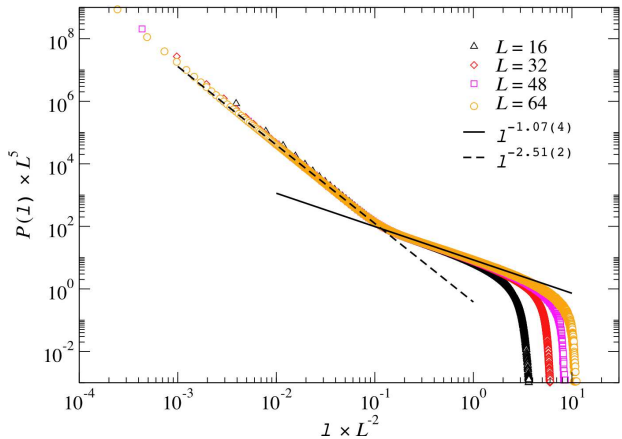}
  \caption{
    (Color online) Probability $P(l)$ of observing a loop of length $l$ in the transition graph between two NN-VB states when sampling the function Eq.~\eqref{eq:psi} on
                              the cubic lattice, for various system sizes $L$. Two regimes can be distinguished for short and long loops with power-law decays with,
                              respectively, best-fit exponents $\alpha=2.51(2)$ and $\beta=1.07(4)$ (lines), in agreement \cite{note_scaling} with the scaling found in
                              Ref.~\cite{jaubert:11} for a spin ice system. Error bars are smaller than symbols.
  }
  \label{fig:Pl}
  \end{center}
\end{figure}

Further insight into the occurrence of AF order is given by the analysis of the length $l$ of loops appearing in the transition graphs ${\langle c_{1} | c_{2} \rangle}$. Indeed, the presence of long-range AF correlations requires long, system-spanning, loops with a size distribution that falls ``slowly
enough"~\cite{beach:06}. We keep record of loop sizes during the simulations, and plot in Fig.~\ref{fig:Pl} the probability $P(l)$ for the appearance of a loop of length
$l$ for the state Eq.~\eqref{eq:psi} on the cubic lattice. Intriguingly, $P(l)$ displays a two-regime
behavior remarkably similar to what was found for `Dirac strings' -- the flux loops of the emergent gauge field of the Coulomb
phase -- in spin-ice~\cite{jaubert:11}: $P(l)\sim l^{-\alpha}$ for short and $P(l)\sim l^{-\beta}$
for long loops, with best-fit exponents $\alpha=2.51(2)$ and $\beta=1.07(4)$. We note that these are the exponents characteristic of random walks \cite{jaubert:11}.
The naive expectation for the statistics of the nontrivial loops in the transition graph generated by the superposition of different VB coverings would be that of a
self-avoiding but otherwise random walk. Random due to the fact that correlations between VBs in the same covering decay quickly (like $r^{-3}$), and self-avoiding
because the loops cannot self-intersect. However, as it is also the case for dense polymer solutions \cite{degennes:79}, the need to avoid all the other loops in the
transition graph counterbalances self-avoidance, with the net effect of producing exponents characteristic of the Brownian random walk.

We finally remark that the long-range tail of $P(l)$, slowly falling as $\sim l^{-1}$, ensures long-range AF correlations~\cite{beach:06} and is caused by winding loops in the
transition graphs.

{\it Four-Spin Correlations --- }
We proceed now to the characterization of `dimer-dimer" four-spin connected correlations,
\begin{equation}
C_{ijkl}= \langle ({\mathbf S}_{i}\cdot {\mathbf S}_{j}) ({\mathbf S}_{k}\cdot {\mathbf S}_{l}) \rangle -\langle {\mathbf S}_{i}\cdot
{\mathbf S}_{j} \rangle \langle {\mathbf S}_{k}\cdot {\mathbf S}_{l}\rangle~,
\label{eq:4point}
\end{equation}
where each pair of indexes $i$, $j$ and $k$, $l$ denote NN sites (dimers). Anticipating the dipolar structure of these
correlators, we evaluate $r^{3} \sigma C_{ijkl}$ \cite{huse:03}, where $r$ is the distance between dimers and $\sigma = +1$ ($-1$) if the sites $i$ and $k$
lie on the same (different) sublattice(s), along selected directions highlighted in Fig.~\ref{fig:lattices}.

\begin{figure}
\begin{center}
  \includegraphics*[width=0.28\textwidth,angle=270]{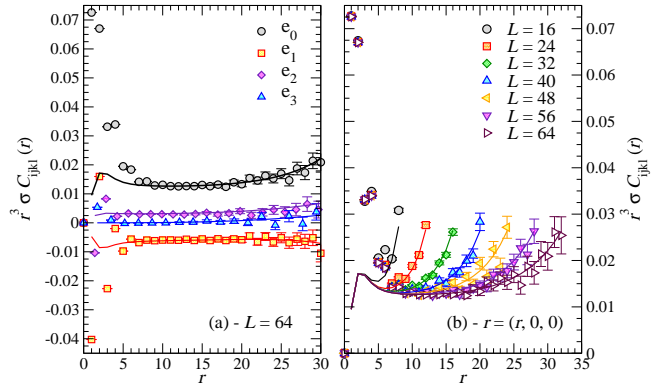}
  \caption{
    (Color online) Four-point connected correlations [Eq.~\eqref{eq:4point}] for the NN-VB state on the cubic lattice, multiplied by the sublattice
                              sign factor $\sigma$
                              and the cube of the distance between dimers, $r^{3}$, for: (a) $L=64$ and specified directions and (b) various system
                              sizes and fixed direction ${\mathbf{e}}_0$ [special directions are highlighted in Fig.~\ref{fig:lattices}(a)]. In both panels,
                              symbols indicate data from MC simulations, and lines correspond to the finite size expression Eq.~\eqref{eq:FSizeCube}.
  }
  \label{fig:dipolar_cube}
  \end{center}
\end{figure}

For the NN-VB wave function on the cubic lattice, we plot $r^{3} \sigma C_{ijkl}$ {\em versus} $r$ for $L=64$ in Fig.~\ref{fig:dipolar_cube}(a), along the same
directions [parallel to the vectors ${\bf e}_{a=0,1,2,3}$ in Fig.~\ref{fig:lattices}(a)] considered in the analysis of classical dimers in Ref.~\onlinecite{huse:03}.
Curves for all directions tend to saturate, evidencing a dipolar structure
\begin{equation} 
C_{a}(r)=  \frac{3 \cos^2(\theta_{a}) - 1}{K r^3}~,
\label{eq:dipolar}
\end{equation}
where $K$ is a constant and $\theta_{a}$ is the angle between the reference dimer along ${\bf e}_0$ and ${\bf e}_{a=0,1,2,3}$.
Finite-size effects are accounted for by the lattice sum expression derived in Ref.~\onlinecite{huse:03}:
\begin{equation}
\sigma C_{ijkl}= \frac{1}{KL^{3}} \left[ 1 + \sum_{{\mathbf{k}} \neq {\mathbf{0}}} \frac{2 - \cos k_y - \cos k_z}{3 -  \sum_i \cos k_i}
                              e^{i {\mathbf{k}} \cdot {\mathbf{r}} } \right]~,
\label{eq:FSizeCube}
\end{equation}
where $k_{i=x,y,z}$ are the reciprocal vectors of the simple cubic lattice. Finite size results obtained from Eq.~\eqref{eq:FSizeCube}, with the single adjustable
parameter set to $K=13.0(5)$, are depicted as solid lines in Fig.~\ref{fig:dipolar_cube} and are in excellent agreement with the MC data (symbols) for all system
sizes and directions, further establishing the dipolar structure of four-spin correlations for Eq.~\eqref{eq:psi} on the cubic lattice.

\begin{figure}
\begin{center}
  \includegraphics*[width=0.28\textwidth,angle=270]{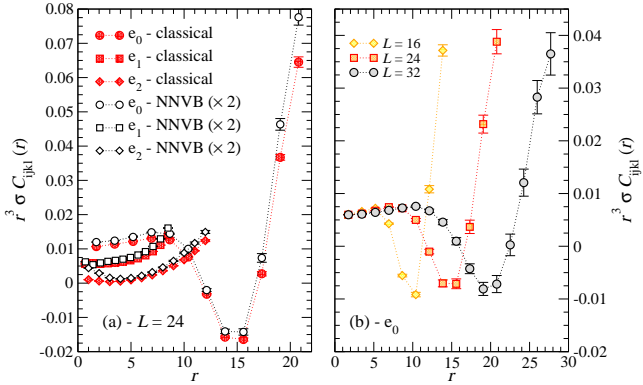}
  \caption{
    (Color online) MC data for $r^{3} \sigma C_{ijkl}$ [Eq.~\eqref{eq:4point}] for the NN-VB state on the diamond lattice as a function of $r$ along
                              special directions indicated in Fig.~\ref{fig:lattices}(b). (a) Data for $L=24$ and specified directions for {\em classical dimers} (filled
                              symbols) and the NN-VB wave function (open symbols; data multiplied by $2$).
                              (b) Correlations for the NN-VB wave function along the direction ${\mathbf{e}}_0$ [see Fig.~\ref{fig:lattices}(b)] for various
                              system sizes. Lines are guides to the eye.
  }
  \label{fig:dipolar_diamond}
  \end{center}
\end{figure}

We have also computed $r^{3} \sigma C_{ijkl}$ for Eq.~\eqref{eq:psi} on the diamond lattice, along the directions ${\mathbf{e}}_a$ ($a=0,1,2$) indicated in
Fig.~\ref{fig:lattices}(b). Fig.~\ref{fig:dipolar_diamond}(a) shows that, similarly to what happens in the case of the cubic lattice, four-point correlations
are also consistent with a dipolar structure [Eq.~\eqref{eq:dipolar}] for the diamond lattice, although finite size effects are stronger and saturation at large $r$ less evident;
specially along ${\mathbf{e}}_0$, as shown in Fig.~\ref{fig:dipolar_diamond}(b). We have also performed MC simulations for {\em classical dimers} on the same
geometry, using the algorithm of Ref.~\onlinecite{sandvik:06}, and find an essentially identical, and equally strong, finite size dependence for
dimer-dimer correlations also in this case [filled symbols in Fig.~\ref{fig:dipolar_diamond}(a)]. We have not been able to derive a finite-size expression similar
to Eq.~\eqref{eq:FSizeCube} for the diamond lattice, even though we suspect this is possible.

{\it Discussion --- }
In conclusion, large-scale MC simulations of NN-VB wave functions on the cubic and diamond lattices have allowed us to obtain (numerically) {\em exact} results for such quantum states, a rare situation for quantum magnetism in 3d. Remarkably, despite of the fact that NN-VB states were originally introduced to mimic spin liquid states in 2d, their versions on bipartite 3d lattices are magnetically ordered. As a consequence, variational approaches relying on NN-VB states, that are only able to describe magnetically disordered states in 2d \cite{mambrini:00,mambrini:06,albuquerque:11b}, may be expected to be useful in analyzing the interplay between disordered and SU($2$)-broken phases on bipartite 3d lattices.

Additionally, we have shown that four-spin correlations possess a dipolar structure, a feature reminiscent of the Coulomb phase of classical dimers \cite{huse:03}. This suggests that the wave functions investigated here may be employed in a variational description of spin-half $U(1)$ quantum spin liquids~\cite{hermele:04}. However,
such an approach is hindered by the presence of magnetic order, that prevails over dipolar correlations. An alternative route involves considering VBs with
enhanced symmetry: for instance, SU($N$) VBs configurations have reduced overlaps as $N$ increases. Correspondingly, the transition graph loops $\langle c_1 | c_2 \rangle$ exhibit an increase in the (bare) loop tension, leading to a geometric transition where the long loops vanish together with magnetic order. Indeed, the large $N$ limit corresponds to orthogonal dimer coverings, for which spin correlations vanish beyond two lattice spacings. 

Finally, we comment on the basic structure of a field-theoretical description of our results, that must account for the presence of two {\em distinct} emergent Coulomb structures. On one hand, four-point dipolar correlations can be expected to be described by a fictitious magnetic field, similar to the case of hard-core dimer configurations~\cite{huse:03}. In 2d, this approach is formulated in terms of a ``dimer-height" (generalizing, in $d=3$, to a ``dimer-gauge'') variable. Such a dimer framework has been put in place by Damle {\it et al.}~\cite{damle:12}, who provide an approximate mapping of the NN-VB problem to a {\it classical} dimer model with {\it interactions} -- the possibility of a dimer phenomenology having been noted in Ref.~\onlinecite{tang:11} (see also Ref.~\onlinecite{cano:10}). Combined with the known phase diagram~\cite{alet:06} of the corresponding interacting dimer model on the cubic lattice, the system is predicted to stay in the Coulomb phase with dipolar dimer correlations [Eq.~\eqref{eq:dipolar}] as observed. On the other hand, this {\em cannot} also account for AF long-range order, as this order can disappear independently of the presence of the algebraic dimer correlations. One therefore requires a second set of ``loop-gauge" variables, whose Coulomb phase captures the loop size distribution of Fig.~\ref{fig:Pl}. We hope that these remarks, together with our numerical results, may stimulate field-theory work to describe the unusual properties of the NN-VB wavefunctions reported herein.

\begin{acknowledgments}
We thank K.~Damle for a critical reading of the manuscript, and L.~Jaubert and A.~W.~Sandvik for very useful discussions. Our MC codes are based upon the ALPS libraries \cite{troyer:98,albuquerque:07}. This work was performed using HPC resources from GENCI-CCRT, GENCI-IDRIS (grants 2010-x2010050225 and 2011-x2011050225) and CALMIP and is supported by the French ANR program ANR-08-JCJC-0056-01, ANR-11-IS04-005-01, and the Indo-French Centre for the Promotion of Advanced Research (IFCPAR/CEFIPRA) under Project 4504-1. A.~F.~A.~acknowledges financial support from Faperj (Brazil).
\end{acknowledgments}

\bibliographystyle{apsrev}

\begin{thebibliography}{46}
\bibitem[{not()}]{note_scaling}
\bibinfo{note}{Notice, however, that we normalize as $\sum_l P(l)=1$, whereas
  Ref.~\onlinecite{jaubert:11} uses $\sum_l P(l)=16L^3$. This difference
  accounts for the extra factor $L^3$ (vertical axis) in the present case.}
\end{thebibliography}

\end{document}